%% file: ecosmo01.tex
\documentclass[12pt]{article}
\usepackage{graphicx}

\newsavebox{\sboxpubnumber}
\newsavebox{\sboxpubdate}
\newcommand{\pubdate}[1]{\begin{lrbox}{\sboxpubdate}{#1}\end{lrbox}}
\newcommand{\pubnumber}[1]{\begin{lrbox}{\sboxpubnumber}{\begin{tabular}{l} #1 \\
				 \usebox{\sboxpubdate}
				 \end{tabular}}
                           \end{lrbox}
                           \pubblock}
\newcommand{\Title}[1]{\begin{center} {\Large #1 } \end{center}}
\newcommand{\Author}[1]{\begin{center}{ \sc #1} \end{center}}
\newcommand{\Address}[1]{\begin{center}{ \it #1} \end{center}}

\newcommand{\pubblock}{\rightline{
			\usebox{\sboxpubnumber}}}
\newenvironment{Abstract}{\begin{quotation}  }{\end{quotation}}
\newenvironment{Presented}{\begin{quotation} \begin{center}
             PRESENTED AT\end{center}\bigskip
      \begin{center}\begin{large}}{\end{large}\end{center}
      \end{quotation}}
\newcommand{\Acknowledgements}{\bigskip  \bigskip \begin{center} \begin{large}
             \bf ACKNOWLEDGEMENTS \end{large}\end{center}}

\input econfmacros.tex
\def\lsi{\raise0.3ex\hbox{$<$\kern-0.75em\raise-1.1ex\hbox{$\sim$}}}
\def\gsi{\raise0.3ex\hbox{$>$\kern-0.75em\raise-1.1ex\hbox{$\sim$}}}
\newcommand{\lsim}{\mathop{\lsi}}
\newcommand{\gsim}{\mathop{\gsi}}

\begin{document}

\begin{titlepage}
\pubdate{15 September, 2001}                    
\pubnumber{DAMTP-2001-102 \\ hep-ph/0111200} 

\vfill
\Title{Baryogenesis at the End of Hybrid Inflation}
\vfill
\Author{Arttu Rajantie}
\Address{DAMTP, CMS\\
University of Cambridge\\
Wilberforce Road\\
Cambridge CB3 0WA\\
United Kingdom}
\vfill
\begin{Abstract}
The baryon asymmetry of the universe may originate in the
phase transition at the end of hybrid inflation, provided that
the reheat temperature is low enough.
I show that if the field that triggers the end of inflation is the
electroweak Higgs field and CP is violated, 
the transition leads to baryon asymmetry
even if no preheating or non-thermal symmetry restoration takes place.
I estimate the strength of this effect and the constraints
it imposes on the inflationary model.
\end{Abstract}
\vfill
\begin{Presented}
    COSMO-01 \\
    Rovaniemi, Finland, \\
    August 29 -- September 4, 2001
\end{Presented}
\vfill
\end{titlepage}
\def\thefootnote{\fnsymbol{footnote}}
\setcounter{footnote}{0}

\section{Introduction}
In order to explain the baryon asymmetry of the universe, a theory
must satisfy three conditions~\cite{Sakharov:1967dj}: 
It must (obviously) violate baryon
number, but it must also violate the C and CP symmetries, and at the
same time the baryon-number violating interaction must be out of
thermal equilibrium.
As the CP violation present in the standard model of particle physics
is not strong enough, any proposal must include some extra source of
CP violation, but apart from that, the standard model seems to have
all the requires properties~\cite{Rubakov:1996vz}. 
At high temperatures, the baryon number
is violated by sphaleron processes~\cite{Kuzmin:1985mm}, 
and the electroweak phase
transition gives rise to the necessary non-equilibrium state. 

Unfortunately, creating the baryon asymmetry is not enough: It
must also survive until the present day, and therefore the baryon
number violation must stop before the fields equilibrate.
In the standard scenario of electroweak baryogenesis, a
strongly first-order phase transition is needed to avoid this baryon
washout. 
More precisely, the jump of the Higgs
expectation value should be of the same order as the critical 
temperature~\cite{Shaposhnikov:1986jp},
and numerical simulations~\cite{Kajantie:1997qd} 
have shown that this is not the case
for any realistic Higgs mass.

This problem can, however, be avoided if the energy scale of inflation
is well below the electroweak 
scale~\cite{Krauss:1999ng,Garcia-Bellido:1999sv}, because then the equilibrium
temperature of the universe never exceeds the electroweak critical
temperature, no sphaleron processes occur and no baryon washout takes
place. The problem, then, is how to create the baryon asymmetry in the
first place.

In Refs.~\cite{Krauss:1999ng,Garcia-Bellido:1999sv}, 
it was argued that this is possible in hybrid models
of inflation, if the inflaton field resonates with the electroweak
Higgs field. This preheating heats up the long-wavelength modes of the
Higgs and gauge fields, and leads to baryon number violation, which
stops as soon as the fields start to approach thermal equilibrium.
For this mechanism to work, both the preheating and the baryon number
violation must be extremely efficient, but numerical
simulations~\cite{Rajantie:2001nj} seem 
to indicate that it is very difficult to
generate enough baryons.

On the other hand, it was recently shown in Ref.~\cite{Copeland:2001qw} 
that even without any preheating at all,
electroweak-scale hybrid inflation generally leads to baryogenesis,
as the breakdown of the SU(2) gauge invariance leads to a non-zero
Higgs winding number, which decays into baryons.
In this talk, I will review this argument and estimate the amount of
baryon asymmetry generated by this mechanism.

\section{Hybrid inflation}
In models of hybrid inflation~\cite{Linde:1994cn}, 
the inflaton field $\sigma$ is coupled to
another scalar field $\phi$, which becomes unstable at a certain critical
value of $\sigma_c$. This causes the slow roll conditions to break
down, and inflation ends. The simplest realization of this idea is
the potential
\begin{equation}
\label{equ:hybrid_pot}
V(\sigma,\phi)=\frac{1}{2}m_\sigma^2\sigma^2+g^2\sigma^2\phi^\dagger\phi
-|m_\phi^2|\phi^\dagger\phi+\lambda\left(\phi^\dagger\phi\right)^2
+\frac{m_\phi^4}{4\lambda}.
\end{equation}
In the following, we will assume that $\phi$ is the electroweak Higgs
field.
The amplitude of the CMB fluctuations forces $m_\sigma$ to be
extremely small, around $10^{-10}~{\rm eV}$, and for our purposes, we
can treat it as zero. 

During inflation, the inflaton $\sigma$ has a large value,
$\sigma\gg\sigma_c=m_\phi/g$. Therefore the effective mass term of the
Higgs field, 
\beq
m_\phi^2(\sigma)=-|m_\phi^2|+g^2\sigma^2=g^2(\sigma^2-\sigma_c^2),
\eeq{equ:effmass} 
is positive
and the electroweak SU(2) symmetry is restored.
The inflaton $\sigma$ slowly rolls down the potential, and
when it reaches $\sigma_c$, 
$m_\phi^2(\sigma)$ becomes negative, implying that the $\phi$ field
becomes unstable and the symmetry gets spontaneously broken.

In order to avoid the baryon washout, the final reheat temperature
$T_{\rm rh}$
must be far enough below the electroweak critical 
temperature~\cite{Shaposhnikov:1986jp},
$
T_{\rm rh}\lsim 150~{\rm GeV},
$
and this constrains the potential at $\sigma=\sigma_c$ to be below the
corresponding critical energy density,
\beq
V(\sigma_c,0)\approx 
\rho(T_{\rm rh})\approx \frac{\pi^2}{30}g_*T_{\rm rh}^4\lsim
10^{10}~{\rm GeV^4},
\eeq{equ:potbound}
where $g_*\approx 100$ is the effective number of degrees of freedom at the
electroweak scale.
This implies that the Hubble rate is around $10^{-5}~{\rm eV}$, well below any
interesting energy scale, and therefore we can ignore the expansion of
the universe altogether.

It is actually rather natural to identify the 
field $\phi$ with the electroweak Higgs
field. Indeed, if it were merely a real scalar field, the symmetry
breakdown would lead to formation of domain walls, with disastrous
consequences. The breakdown of any global continuous symmetry would
imply the existence of Goldstone bosons, which have not been observed,
and therefore the broken symmetry must be a
gauge invariance. Then, if we want to avoid the formation of
monopoles or cosmic strings, the simplest choice is that $\phi$ is an
SU(2) doublet, which means that it is the electroweak Higgs field.

This has the consequence that it fixes the couplings $\lambda$ and
$m_\phi^2$, if we assume that the Higgs mass is $m_{\rm H}\approx 115~{\rm
GeV}$:
\beq
\lambda\approx 0.11,\qquad m_\phi^2=\frac{m_{\rm H}^2}{2}\approx 
-6.6\times 10^3~{\rm GeV}^2.
\eeq{equ:fixparams}
This implies that
\beq
V_0\equiv
V(\sigma_c,0)=V(0,0)=\frac{m_{\rm H}^4}{16\lambda}\approx 10^8~{\rm GeV}^4,
\eeq{equ:fixpot}
which satisfies the bound in Eq.~(\ref{equ:potbound}) easily.
The final reheat temperature is
\beq
T_{\rm rh}\approx \left(
\frac{15}{8\pi^2g_*\lambda}
\right)^{1/4}m_{\rm H}\approx 42~{\rm GeV}.
\eeq{equ:fixTrh}
In this simplest realization, the only remaining free parameter is
$g$.
We shall assume that $g\sim O(1)$, even though this means that
radiative corrections arising from Higgs loops are going to cause
problems. These problems are alleviated in inverted hybrid
models~\cite{Copeland:2001qw}.

\section{Baryogenesis}
When the inflaton field $\sigma$ reaches its critical value
$\sigma_c$, the Higgs field becomes unstable and the SU(2) gauge
invariance breaks spontaneously. Some aspects of this process have
recently been discussed in Refs.~\cite{Felder:2001hj,Felder:2001kt}. 
For our purposes, it is
enough to note that symmetry-breaking phase transitions typically lead
to formation of topological defects via the well-known Kibble-Zurek
mechanism~\cite{Kibble:1976sj,Zurek:1985qw,Rajantie:2001ps}.

Of course, the electroweak theory does not contain any genuine
topological defects. The zeroth, first and second homotopy groups of
the vacuum manifold are trivial, and therefore no domain walls,
strings or monopoles are formed. However, as the vacuum manifold is
a three-sphere, it has a non-trivial third homotopy group, whose
elements are given by the integer-valued 
Higgs winding number $N_{\rm H}$~\cite{Turok:1990in}
\beq
N_{\rm H}=-\frac{1}{24\pi^2}\int d^3x \eps^{ijk}{\rm Tr}\left(
\partial_i\Phi^\dagger \partial_j\Phi \partial_k\Phi^\dagger \Phi
\right).
\eeq{equ:winding}
Here $\Phi$ is a unitary matrix
$\Phi=\hat\phi^4+i\sigma^i\hat{\phi}^i$
with 
$\phi=|\phi|\left(\hat\phi^2+i\hat\phi^1,\hat\phi^4-i\hat\phi^3\right).$

If the SU(2) symmetry were global, the Higgs winding number would
label physically distinct vacua of the theory.
In the gauge theory, however, there is for every Higgs field configuration a
corresponding 
gauge field configuration that compensates exactly for the gradient
energy, and in such a configuration the
Chern-Simons number
\begin{equation}
N_{\rm CS} = {g^2 \over 16 \pi^2} \int d^3 x
\epsilon^{ijk} {\rm Tr} \left(
F_{ij}A_k+\frac{2}{3}ig A_iA_jA_k\right)
\label{cs_number}
\end{equation}
is equal to the Higgs winding number, $N_{\rm H}=N_{\rm CS}$.
Large gauge transformations can change both $N_{\rm CS}$ and $N_{\rm H}$,
and therefore a vacuum with equal but non-zero Higgs winding
and Chern-Simons numbers is always merely a gauge transform of the
vacuum with $N_{\rm H}=N_{\rm CS}=0$, and therefore physically
indistinguishable from it.

Or, more precisely, it would be if there were no fermions. 
An anomaly relates the
Chern-Simons number to the baryon and lepton numbers,
and in particular, 
\begin{equation}
\Delta N_B=3\Delta N_{\rm CS}.
\end{equation}
This means that if an instanton or a sphaleron transition moves the
system from one Chern-Simons vacuum to another, the baryon number
changes as well. This is the basic mechanism of electroweak
baryogenesis. Starting from a state with $N_B=0$, some non-equilibrium
mechanism changes the Chern-Simons number, and when the universe
finally settles down in that vacuum, all other physical observables
are identical to the original vacuum, but the baryon and lepton
numbers have changed.

In the standard scenario~\cite{Rubakov:1996vz}, 
the baryon number violation takes
place on the bubble walls during a first-order electroweak phase
transition, but lattice simulations show that in the standard model,
the transition is not of first order. In a modified scenario,
preheating after electroweak-scale inflation leads to non-thermal
symmetry restoration, which allows the Chern-Simons number to
change~\cite{Krauss:1999ng,Garcia-Bellido:1999sv}. 
However, that scenario is very sensitive to the
non-equilibrium dynamics during and after preheating, and numerical
simulations indicate that the baryon number violation may not be as
strong as was hoped~\cite{Rajantie:2001nj}.

However, the Chern-Simons number can also
change if it simply happens to differ from the Higgs
winding number~\cite{Turok:1990in,Turok:1991zg}, 
and this does not need high temperatures or
symmetry restoration. The fields simply relax to the
vacuum configuration with $N_{\rm H}=N_{\rm CS}$, and this can happen in two
different ways: Either the Higgs unwinds
as in the global theory, in which case $N_{\rm H}$ changes, or the
Chern-Simons number of the gauge field changes. In the latter case,
baryons are created.
The presence of CP violation biases this 
process~\cite{Turok:1990in,Turok:1991zg}, so that even if the
Higgs winding number averages to zero and only has spatial
fluctuations, the regions with, say, positive $N_{\rm H}$ are more likely
to form baryons than those with negative $N_{\rm H}$ are to form
antibaryons. This leads to baryon asymmetry even if the mechanism which
forms the Higgs winding is symmetric.

\begin{figure}[htb]
    \centering
    \includegraphics[height=1.5in]{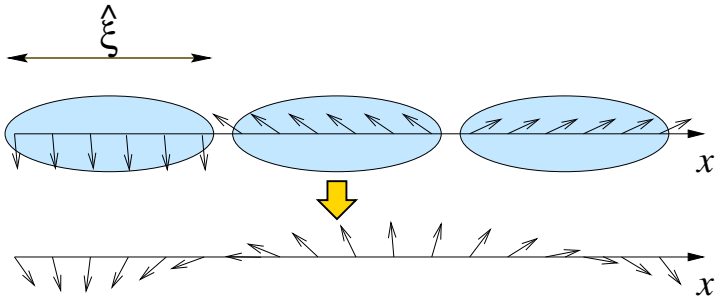}
    
\caption{The Kibble mechanism in a one-dimensional complex field
theory. After the transition, the Higgs field (arrows) is correlated
only inside domains of size $\hat{\xi}$, and this leads to the
formation of Higgs windings.
}
    \label{fig:winding}
\end{figure}

In our case~\cite{Copeland:2001qw}, 
these spatial variations in the Higgs winding number are
generated by the Kibble-Zurek mechanism~\cite{Kibble:1976sj,Zurek:1985qw}.
At the end of inflation, the universe is cold, and we can ignore the
gauge field~\cite{Hindmarsh:2000kd,Rajantie:2001ps}. In the gauge
where $\vec{A}=0$, the direction of the Higgs field has a well defined
meaning, and in the phase transition spatial domains will form inside
each of which the direction will be roughly constant, but between
which it is totally uncorrelated. However, the Higgs field must
smoothly interpolate between these domains, and as illustrated by the
one-dimensional example in Fig.~\ref{fig:winding}, this generally leads to
non-zero Higgs winding number. Because the only relevant length scale
for this process is the domain size $\hat{\xi}$, we can estimate that
the typical Higgs winding density after the transition is
\beq
n_{\rm H}=N_{\rm H}/V\sim \hat{\xi}^{-3}.
\eeq{equ:nHpred}
Positive and negative values of $n_{\rm H}$ are obviously just as likely,
and therefore Eq.~(\ref{equ:nHpred}) should be interpreted as the rms
value
$n_{\rm H}^{\rm rms}=\langle n_{\rm H}^2\rangle^{1/2}.$

Assuming that each positive Higgs winding has a certain probability
$p_+=p_0+\eps$ to decay by changing $N_{\rm CS}$ and thereby
creating 
baryons,
and correspondingly that the decay of a negative winding to
antibaryons has a lower probability $p_-=p_0-\eps$, we obtain the estimate
\beq
n_B\approx 3\eps n_{\rm H}\sim \eps\hat{\xi}^{-3},
\eeq{equ:nBestimate}
where $\eps$ depends on the strength of the CP violation.

When the fields finally thermalize, they reach the reheat temperature
in Eq.~(\ref{equ:fixTrh}), which corresponds to the entropy density
$s=(2\pi^2/45)g_*T_{\rm rh}^3\sim 2.1 m_{\rm H}^3.$
In order to explain the observed baryon asymmetry $n_B/s\sim
3\times 10^{-10}$, we therefore need
\beq
n_{\rm H}\gsim 10^{-9}\eps^{-1}m_{\rm H}^3.
\eeq{equ:reqnH}

\section{Estimating $n_{\rm H}$}
Although defect formation is usually discussed in the context of
thermal phase transitions~\cite{Kibble:1976sj,Zurek:1985qw,Rajantie:2001ps}, 
the same principles apply to our
case. The role of the temperature is played by the inflaton field,
which gives the Higgs field a time-dependent mass term
(\ref{equ:effmass}). Near $\sigma_c$, we can approximate
\beq
m_\phi^2(t)=g^2(\sigma(t)^2-\sigma_c^2)\approx
-2g^2\sigma_c|\dot{\sigma}|t+O(t^2).
\eeq{equ:linmass}
The correlation length of the Higgs field is given by the inverse mass
\beq
\xi(t)\approx\left(2g^2\sigma_c|\dot{\sigma}||t|\right)^{-1/2},
\eeq{equ:corrlength}
which shows that the correlation length diverges at the critical point.
In the absence of any thermal fluctuations, this is equal to the
relaxation time of the field, $\tau(t)=\xi(t)$,
and so the dynamics of the field also gets slower. Because of this
critical slowing down, a point is eventually reached when the field
cannot respond to the change of the mass, and therefore, instead of
actually diverging, the correlation length reaches a maximum value
$\hat{\xi}$, which determines the size of the correlated domains and
thereby also the Higgs winding density~(\ref{equ:nHpred}).

The value of $\hat{\xi}$ is approximately given by the correlation
length at the time when $|t|$ is equal to the relaxation time, which
gives
\beq
\hat{\xi}=\xi\left(t\!=\!-\hat{\xi}\right)
\approx \left(2g^2\sigma_c|\dot\sigma|
\right)^{-1/3},
\eeq{equ:esthatxi}
or
\beq
n_{\rm H}\sim 2g^2\sigma_c|\dot\sigma|\approx gm_{\rm H}|\dot\sigma|.
\eeq{equ:estnH}

If we use the slow roll condition
\beq
\dot\sigma=-\frac{V'(\sigma)}{3H}
\eeq{equ:slowroll}
together with the potential (\ref{equ:hybrid_pot}), we find that
$\dot\sigma$ is far too low, and therefore we must assume that before
$\sigma_c$, the potential becomes steeper. This may happen because
of radiative corrections, which become much more important when the
effective mass of the Higgs field is low, or it can even be the case
at tree level in more complicated models~\cite{Copeland:2001qw}. 
If we simply require
that the inflation has not ended before $\sigma_c$, we have
\beq
|\dot\sigma|\lsim V_0^{-1/2}\approx \frac{m_{\rm H}^2}{4\sqrt{\lambda}},
\eeq{equ:dotinfla}
which implies
\beq
n_{\rm H}\lsim \frac{g}{4\sqrt{\lambda}}m_{\rm H}^3\approx gm_{\rm H}^3.
\eeq{equ:nHest}
Combining this with Eq.~(\ref{equ:reqnH}) gives the requirement
\beq
\eps g\gsim 10^{-9},
\eeq{equ:epsgbound}
which means that the coupling $g$ and the CP violation $\eps$ do not
have to be particularly strong for this mechanism to be able to
explain the observed baryon asymmetry of the universe.

\section{Conclusions}
In this talk, 
I have discussed the possibility of baryogenesis at the end of 
electroweak-scale hybrid inflation.
I have shown that the Higgs winding number generated by
the Kibble-Zurek mechanism leads to baryon asymmetry, provided that 
CP is violated.
There are
still certain questions that need to be addressed before a concrete
estimate of the resulting baryon density can be obtained:
We need to establish the constant of proportionality in
Eq.~(\ref{equ:nHpred}),
and the dependence of $\eps$ in Eq.~(\ref{equ:nBestimate}) on the CP
violating couplings in the Lagrangian.
Furthermore, we do not yet have a natural model of electroweak-scale
inflation that would have all the necessary properties.

Nevertheless, the estimates I have presented here indicate that in
a simple toy model, it is possible to achieve high enough baryon
densities to explain the observed baryon asymmetry of the universe,
without having to assume that either the CP violation or the coupling
between the Higgs and the inflaton is particularly strong.

\Acknowledgements
I would like to thank Ed Copeland, David Lyth and Mark Trodden for
collaboration on this topic.

\end{document}

%% file: econfmacros.tex
\def\beq{\begin{equation}}
\def\eeq#1{\label{#1}\end{equation}}
\def\eeqn{\end{equation}}

\def\beqa{\begin{eqnarray}}
\def\eeqa#1{\label{#1}\end{eqnarray}}
\def\eeqan{\end{eqnarray}}

\let\bar=\overbar

\def\Dslash{\not{\hbox{\kern-4pt $D$}}}
\def\dslash{\not{\hbox{\kern-2pt $\del$}}}

\def\msb{{\bar{\ssstyle M \kern -1pt S}}}

\def\eps{\epsilon}




%% file: ecosmo01.bbl
\begin{thebibliography}{99}


\bibitem{Sakharov:1967dj}
A.~D.~Sakharov,
Pisma Zh.\ Eksp.\ Teor.\ Fiz.\  {\bf 5} (1967) 32
[JETP Lett.\  {\bf 5} (1967) 24].

\bibitem{Rubakov:1996vz}
V.~A.~Rubakov and M.~E.~Shaposhnikov,
Usp.\ Fiz.\ Nauk {\bf 166} (1996) 493
[Phys.\ Usp.\  {\bf 39} (1996) 461]
[arXiv:hep-ph/9603208].

\bibitem{Kuzmin:1985mm}
V.~A.~Kuzmin, V.~A.~Rubakov and M.~E.~Shaposhnikov,
Phys.\ Lett.\ B {\bf 155} (1985) 36.

\bibitem{Shaposhnikov:1986jp}
M.~E.~Shaposhnikov,
JETP Lett.\  {\bf 44} (1986) 465
[Pisma Zh.\ Eksp.\ Teor.\ Fiz.\  {\bf 44} (1986) 364].



\bibitem{Kajantie:1997qd}
K.~Kajantie, M.~Laine, K.~Rummukainen and M.~E.~Shaposhnikov,
Nucl.\ Phys.\ B {\bf 493} (1997) 413
[arXiv:hep-lat/9612006].

\bibitem{Krauss:1999ng}
L.~M.~Krauss and M.~Trodden,
Phys.\ Rev.\ Lett.\  {\bf 83} (1999) 1502
[arXiv:hep-ph/9902420].

\bibitem{Garcia-Bellido:1999sv}
J.~Garcia-Bellido, D.~Y.~Grigoriev, A.~Kusenko and M.~E.~Shaposhnikov,
Phys.\ Rev.\ D {\bf 60} (1999) 123504
[arXiv:hep-ph/9902449].

\bibitem{Rajantie:2001nj}
A.~Rajantie, P.~M.~Saffin and E.~J.~Copeland,
Phys.\ Rev.\ D {\bf 63} (2001) 123512
[arXiv:hep-ph/0012097].

\bibitem{Copeland:2001qw}
E.~J.~Copeland, D.~Lyth, A.~Rajantie and M.~Trodden,
Phys.\ Rev.\ D {\bf 64} (2001) 043506
[arXiv:hep-ph/0103231].

\bibitem{Linde:1994cn}
A.~D.~Linde,
Phys.\ Rev.\ D {\bf 49} (1994) 748
[arXiv:astro-ph/9307002].



\bibitem{Felder:2001hj}
G.~N.~Felder, J.~Garcia-Bellido, P.~B.~Greene, L.~Kofman, A.~D.~Linde and I.~Tkachev,
Phys.\ Rev.\ Lett.\  {\bf 87} (2001) 011601
[arXiv:hep-ph/0012142].

\bibitem{Felder:2001kt}
G.~N.~Felder, L.~Kofman and A.~D.~Linde,
arXiv:hep-th/0106179.

\bibitem{Kibble:1976sj}
T.~W.~B.~Kibble,
J.\ Phys.\ A {\bf 9} (1976) 1387.

\bibitem{Zurek:1985qw}
W.~H.~Zurek,
Nature {\bf 317} (1985) 505.

\bibitem{Rajantie:2001ps}
A.~Rajantie,
arXiv:hep-ph/0108159.



\bibitem{Turok:1990in}
N.~Turok and J.~Zadrozny,
Phys.\ Rev.\ Lett.\  {\bf 65} (1990) 2331.

\bibitem{Turok:1991zg}
N.~Turok and J.~Zadrozny,
Nucl.\ Phys.\ B {\bf 358} (1991) 471.

\bibitem{Hindmarsh:2000kd}
M.~Hindmarsh and A.~Rajantie,
Phys.\ Rev.\ Lett.\  {\bf 85} (2000) 4660
[arXiv:cond-mat/0007361].

\end{thebibliography}
